\documentclass[aps,prb, twocolumn, groupedaddress, amssymb, amsmath,nofootinbib]{revtex4-2}
\usepackage{bm}
\usepackage{graphicx}
\usepackage{float}
\usepackage{todonotes}
\usepackage{physics}
\usepackage{ulem}

\begin{document}


\title{Quasiparticle focusing of bound states in two-dimensional $s$-wave superconductors}


\author{Mateo Uldemolins}
\author{Andrej Mesaros}
\author{Pascal Simon}

\email{pascal.simon@universite-paris-saclay.fr}
\affiliation{Universit\'e Paris-Saclay, CNRS, Laboratoire de Physique des Solides, 91405, Orsay, France}

\date{\today}

\begin{abstract}
A magnetic impurity on a superconducting substrate induces in-gap Yu-Shiba-Rusinov (YSR)
bound states, whose intricate spatial structure crucially influences the possibilities of engineering
collective impurity states. By means of a saddle-point approximation we study the scattering processes giving rise to YSR states in gapped, two-dimensional superconductors. Further, we develop a theory which relates through a simple analytical expression an arbitrary energy dispersion of normal electrons in a two-dimensional host to the spatial features of the YSR states. Namely, we find that flatter segments of the Fermi surface with large Fermi velocity enhance the local density of states (LDOS) around the impurity. 
Our analytical approximation is quantitatively accurate against tight-binding calculations on various lattices with different Fermi surfaces, and it allows to predict the shape and orientation of YSR states observed in scanning tunneling spectroscopy experiments. We illustrate our results with a model of NbSe\textsubscript{2}. 

\end{abstract}


\maketitle


\section{Introduction}
\label{sec:intro}
The impurity problem has been in the research community spotlight since the dawn of solid state physics. Aside from its fundamental interest, studying the system's response to an impurity, and more generally, to any kind of defect, constitutes a powerful tool to probe the substrate. A notable example is the seminal work of Weissmann \textit{et al.} \cite{weissmann2009realspaceFS} who imaged the Fermi surface (FS) of the host by analyzing scanning tunneling microscopy (STM) topographies around the impurity. Indeed, the local density of states (LDOS) is focused along perpendicular directions to flat sections of the FS, thereby establishing a direct relationship between the anisotropy of the FS and the system's response. This phenomenon known as quasiparticle focusing has been thoroughly studied in the context of Friedel oscillations in normal metals \cite{lounis2011theory_real_imaging}; however, in spite of being widely accepted that it should also occur in superconducting substrates, a formal treatment is lacking.

Conventional superconductors are largely immune to non-magnetic disorder \cite{ANDERSON195926}, nonetheless, magnetic impurities localize quasiparticle excitations known as Yu-Shiba-Rusinov (YSR) states \cite{Yu1965, shiba:classical_spins, rusinov, sakurai:comments, salkola:magnetic_moments} whose energy lies within the superconducting gap. YSR states contain information about the host, for instance, the properties of its band structure \cite{uldemolins2021effect}, the pairing function \cite{Kaladzhyan2015} or coexisting emergent phases such as charge density waves \cite{franke:ysr_nbse2}. But besides their potential as a probing tool \cite{franke:orbital, choi:magnetic_ordering, huang2020:tunneling, huang2020b, schneider2021:spin_polarization, thupakula2021coherent, Choi:2017, Moca:2008}, arrays of YSR states evidence spectral signatures of Majorana zero modes \cite{nadj-perge:majorana_chain,Ruby2015, Pawlak2016,Yazdani2017,Jeon2017,palacio-morales:majorana_chain, Ruby2017} (see \cite{Yazdani2021} for a review), and hence embody a promising pathway towards the realization of topological superconductivity \cite{Nakosai2013,NP2013,pientka:topo, Pientka2014, Braunecker2013,Klinovaja2013,Vazifeh2013,Kim2014,Heimes2014,Li2014,brydon:topo_chain,Ojanen2015b,Heimes_2015,Braunecker2015,Rontynen2016,Schecter2016,Christen2016,Hoffman2016,Li2016a, Li2016b, Poyhonen2014, Ojanen2015a, Hui2015, Zhang2016, Kalad2017}. Understanding the connection between the Fermi surface of the substrate and the spatial properties of YSR states is therefore a question of both fundamental and practical interest.

YSR states were first observed on a Nb(110) substrate more than two decades ago \cite{yazdani:probing} and since, the field has developed immensely \cite{franke:review_shiba}. Most notably, YSR states have been realized on a monolayer NbSe\textsubscript{2} substrate \cite{menard:coherent}, which on the one hand enhances their spatial extent due to its reduced dimensionality, and on the other hand, imprints a distinctive six-fold symmetry on the LDOS. Subsequent experiments on similar substrates found analogous responses \cite{wiesendanger:focusing, thupakula2021coherent} and the accumulation of the LDOS along preferential directions was ascribed to the quasiparticle focusing effect discussed in the first paragraph. However, unlike the charge-density response in a normal metal which decays algebraically and whose anisotropy can only be encoded in an overall prefactor, YSR states are also endowed with an exponential decay length which also reflects the anisotropy of FS. Very recently, Ortuzar and coworkers \cite{ortuzar2021yushibarusinov} obtained a general integral expression of the Green's function of the substrate by approximating the Fermi contours by regular polygons. However, an explicit description of the quasiparticle focusing effect in superconductors, i.e. the link between simple geometrical features of the exact energy dispersion and the LDOS at the energy of a YSR state, is missing. This is exactly the purpose of the present paper.

To reach that goal, we perform a saddle-point approximation valid at large distances from the impurity, inspired by the treatment of normal metals in Ref.~\cite{lounis2011theory_real_imaging}. We unveil
a simple analytical relationship between, on the one hand, the real-space anisotropy of decay, oscillations and amplitude of YSR states, and on the other hand, the momentum-space anisotropy of the
Fermi surface, Fermi velocity and pairing function of the substrate. Further, we reveal the underlying scattering mechanisms leading to the formation of YSR states. Our analytical calculations are qualitatively consistent with experimental STM measurements on NbSe\textsubscript{2}, and remarkably, they provide a quantitatively accurate description of tight-binding calculations on the same compound. Hence we provide a complete description of the quasiparticle focusing effect in $s$-wave superconductors, thereby bringing forth an analytical tool to predict the shape and orientation of YSR states, and ultimately aid the design of collective impurity states. 

The rest of the paper is organized as follows. In Sec.~\ref{sec:model} we present the model Hamiltonian of a classical spin-impurity on an $s$-wave superconductor. In Sec.~\ref{sec:results} we discuss the implications derived from the saddle-point approximation, namely, the scattering processes (Sec.~\ref{subsec:scatt}) and the interpretation of the critical points in the limit of a small superconducting gap (Sec.~\ref{subsec:small_gap}), and we extend the formalism to $\bm{k}$-dependent, gapped pairing functions (Sec.~\ref{subsec:extended}). Finally, in Sec.~\ref{sec:concl} we summarize the conclusions of our work. In Appendix \ref{app:nm} we extend the results in \cite{lounis2011theory_real_imaging} to a two-dimensional normal metal. We detail the calculations in Appendix \ref{app:calc}, we analyze the interplay between the LDOS prefactor and the decay length in Appendix \ref{app:pref} and we present an example beyond the small-gap approximation in Appendix \ref{app:beyond}.
\section{Model Hamiltonian}
\label{sec:model}
We describe the two-dimensional superconducting substrate at mean field level by the standard BCS Hamiltonian for $s$-wave superconductors,
\begin{equation}
    \label{eq:ham_free}
     H_0 = \sum_{\bm{k}\sigma} \varepsilon_{\bm{k}\sigma} c^\dagger_{\bm{k} \sigma} c_{\bm{k} \sigma} + \sum_{\bm{k}} \Delta_{\bm{k}}  \; c^\dagger_{\bm{k} \uparrow} c^\dagger_{\bm{-k} \downarrow} + \mathrm{h.c.}
\end{equation}
For simplicity we assume that spin-orbit coupling in the substrate is negligible, and therefore that spin is a good quantum number. Nevertheless, if that were not the case Eq.~\eqref{eq:ham_free} would be formally equivalent in a pseudo-spin basis. Further, we will consider a substrate with time-reversal symmetry (TRS), and assume that the energy dispersion of the normal electrons $\varepsilon_{\bm{k}\sigma}$ is spin-independent and even in $\bm{k}$. Finally, let us choose a gauge such that the superconducting parameter is real, and assume it to be $\bm{k}$-independent, $\Delta = \Delta^*$. Since the superconducting substrate has TRS, a non-magnetic potential only does not suffice to induce in-gap states. We will consider a point-like, isotropic, magnetic impurity at $\bm{r} = \bm{0}$, described by the Hamiltonian
\begin{equation}
 \label{eq:ham_imp}
 H_{\mathrm{imp}} = -J \left( c^\dagger_{\bm{r}\uparrow} c_{\bm{r}\uparrow} - c^\dagger_{\bm{r}\downarrow} c_{\bm{r}\downarrow} \right)\delta(\bm{r}),
\end{equation}
where $J$ is the Zeeman splitting between spin-up and spin-down superconducting electrons. We note that a complete description of adsorbed atoms and magnetic molecules typically requires adding a non-magnetic scattering potential to the Hamiltonian. The strength of this potential affects the energy of the YSR state and yields some degree of asymmetry between the in-gap DOS at positive and negative bias, however, it does not alter the fundamental properties of the spatial distribution of the quasiparticle excitations, and therefore we will omit it to simplify matters. Furthermore, we neglect any quantum phenomena associated to the magnetic impurity (e.g. Kondo screening) \cite{zitko:adsorbates, vonoppem:real_metals} and any spatial renormalization of the superconducting gap around the impurity \cite{flatte:local,meng:sc_gap_renormalization}. The Bogoliubov-de Gennes Hamiltonian (BdG) of the system in the Nambu basis $\Psi = (\psi_{\uparrow}, \psi_{\downarrow}, \psi_{\downarrow}^\dagger, -\psi_{\uparrow}^\dagger)^T$ reads
\begin{equation}
\mathcal{H} = \varepsilon_{\bm{k}} \tau_z + \Delta \tau_x - J \sigma_z \delta(\bm{r}).
\label{eq:bdg_ham}
\end{equation}
where $\bm{k}$ and $\bm{r}$ designate the electron's momentum and position, and Pauli matrices $\tau_i$ and $\sigma_i$ act on particle-hole and spin space respectively.

The in-gap contribution to the LDOS due to the impurity is given by 
\begin{equation}
    \delta \rho(\bm{r},E) \sim \operatorname{Tr}\{\operatorname{Im}[\hat{G}_0(\bm{r}, \bm{0};E) \hat{T}(E) \hat{G}_0(\bm{0}, \bm{r};E)]\},
\end{equation}
where
\begin{equation}
\begin{split}
\label{eq:bare_prop_1}
\hat{G}_0&(\bm{r_a}, \bm{r_b};E) = \\
&\int \frac{d\bm{k}}{(2\pi)^2}\frac{e^{i\bm{k}\cdot(\bm{r_a}-\bm{r_b})}}{E^2-\Delta^2-\varepsilon_{\bm{k}}^2}
 \begin{pmatrix}
E + \varepsilon_{\bm{k}} & \Delta \\
\Delta & E - \varepsilon_{\bm{k}}
\end{pmatrix}
\end{split}
\end{equation}
denotes the real-space bare propagator from $\bm{r_a}$ to $\bm{r_b}$ at energy $E$ in particle-hole space, and $\hat{T}(E)$ corresponds to the transfer matrix. Since we assumed that the impurity scattering was fully isotropic, the transfer matrix is momentum-independent, and therefore the spatial structure of the LDOS is encoded in the bare propagator \eqref{eq:bare_prop_1}. This further justifies treating the impurity as a classical spin [Eq.~\eqref{eq:ham_imp}]. We note that it is possible to express the energy of the YSR state $E$ in terms of the system's parameters under certain assumptions about the DOS \cite{uldemolins2021effect}, however, it can be easily calculated for an arbitrary energy dispersion numerically, or measured in an STM experiment \cite{menard:coherent}. Therefore, we will treat it as an independent parameter in the calculation. In the following sections we obtain an approximate expression of the integral in Eq.~\eqref{eq:bare_prop_1} far away from the impurity for an arbitrary anisotropic energy dispersion and we discuss its implications.

\section{Results}
\label{sec:results}
\begin{figure*}
    \centering
    \includegraphics[width=1.0\textwidth]{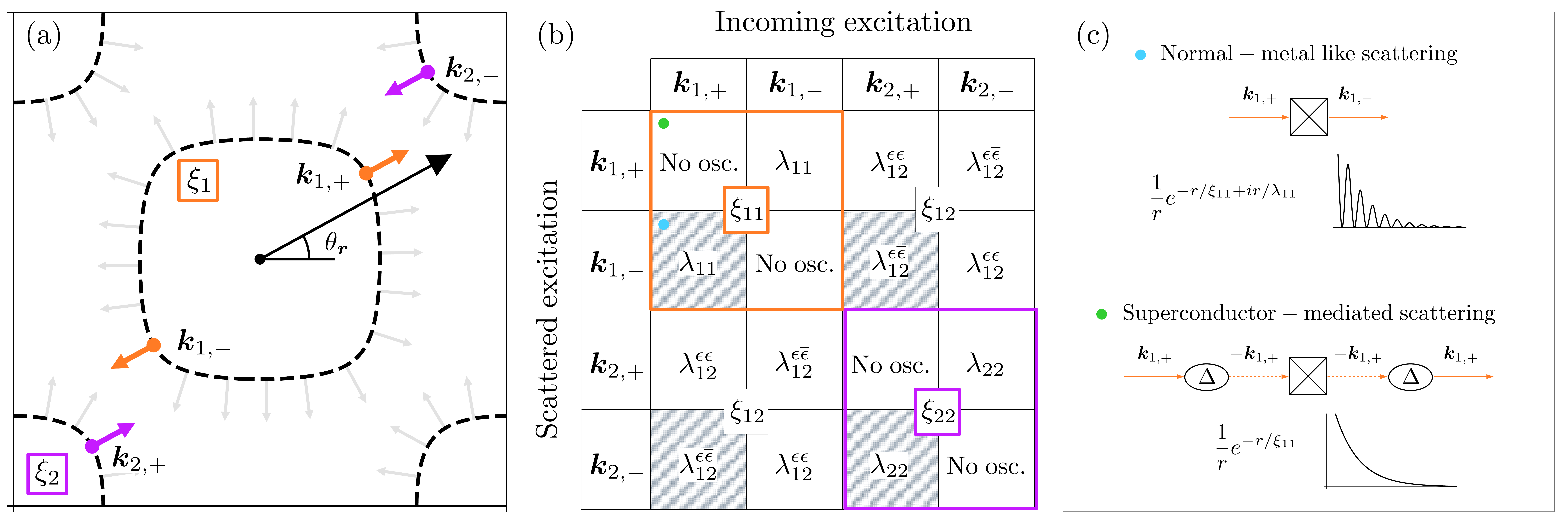}
    \caption{Scattering processes for a Fermi contour with two pockets, $j=2$. (a) The black, dashed lines represent the Fermi contour. The gray arrows indicate the normalized gradient of the energy dispersion at the Fermi contour. The black arrow signals an arbitrary observation angle $\theta_{\bm{r}}$. The color markers indicate the real part of the critical points on the Fermi contours (see Sec.~\ref{subsec:small_gap} for details) for the observation direction $\theta_{\bm{r}}$ and each color corresponds to a different Fermi pocket. The color arrows represent the normalized gradient of the energy dispersion evaluated at the critical points, which is parallel and antiparallel to the observation direction. (b) Summary of the oscillation and decay lengths in the relevant scattering. Colored frames correspond to intrapocket processes. Shaded entries indicate the processes present in a normal metal. Note that we dropped the redundant labels in $\lambda_{j,j'}^{\epsilon, \epsilon'}$ to lighten the notation. (c) Examples of normal-metal like and condensate-mediated scattering process [cf. Eqs.~\eqref{eq:process_ee_2} and ~\eqref{eq:process_eh} respectively]. Color markers indicate the corresponding entry in panel (b).}
    \label{fig:scattering}
\end{figure*}

To calculate $\hat{G}_0(\bm{r},\bm{0};E)$ and $\hat{G}_0(\bm{0},\bm{r};E)$ in the large $r$ regime we start from the idea of the saddle-point approximation technique (see Appendix \ref{app:nm}) and generalize it to the complex plane (see Appendix \ref{app:calc} for details). We assume that $r k_{\mathrm{F, \; min}} \gg 1$ where $k_{\mathrm{F, \; min}}$ is the mimimum Fermi wave vector of an arbitrary Fermi surface. For an isotropic Fermi surface, it boils down to the usual condition $r k_{\mathrm{F}} \gg 1$. We do not treat the superconducting gap self-consistently because the gap should be modified only on a short lengthscale $rk_{\mathrm{F}} \sim 1$ \cite{flatte:local}, much below the lengthscale on which we apply the saddle-point approximation. In essence we replace the integral in momentum space by a sum of the integrand evaluated at the critical points $\bm{k}_j(\theta_{\bm{r}})$ giving the largest contribution to the integral,
\begin{equation}
    \label{eq:approx}
    \hat{G}_0(\bm{r},\bm{0};E) \sim \sum_{j} e^{i\bm{k}_j(\theta_{\bm{r}}) \cdot \bm{r}}G_0[\bm{k}_j(\theta_{\bm{r}});E].
\end{equation}
The set of critical points depends on the observation direction $(\theta_{\bm{r}})$ and they satisfy the following conditions,
\begin{subequations}
\label{eq:crit_sol}
\begin{align}
\varepsilon_{\bm{k}_{j, \pm}(\theta_{\bm{r}})} &= \pm i \omega,\label{eq:crit_sol_energy}\\
\bm{\nabla}\varepsilon_{\bm{k}_{j, \pm}(\theta_{\bm{r}})} &=  \pm|\bm{\nabla}\varepsilon_{\bm{k}_{j, \pm}(\theta_{\bm{r}})}| \hat{\bm{r}},\label{eq:crit_sol_gradient}
\end{align}
\end{subequations}
where $\omega^2 = \Delta^2 - E^2$. To understand the nature of the critical points it is insightful to compare Eqs.~\eqref{eq:crit_sol} with their analogue for a charge impurity embedded in a normal metal. The latter result was originally discussed by Lounis \textit{et al}.~\cite{heinze2000realimaging, lounis2011theory_real_imaging} for three-dimensional systems and we derive its two-dimensional counterpart in Appendix \ref{app:nm}. Similarly to the normal metal scenario, the gradient of the energy dispersion evaluated at the critical points $\bm{k}_{j,+}(\theta_{\bm{r}})$ is also parallel to the observation direction [Eq.~\eqref{eq:crit_sol_gradient}] in the superconductor scenario. Therefore, in both situations disconnected Fermi contours give rise to multiple critical points which we denote with the subscript $j$. Precisely, if the curvature of the Fermi contours is strictly positive which we will assume in the rest of the paper, $j=1,\dots,N$ where $N$ is the number of non-equivalent Fermi pockets in the First Brillouin Zone (FBZ). However, there are two crucial differences:

First, we note that for a given observation direction $\theta_{\bm{r}}$ and a given Fermi pocket $j$ there are two critical points in momentum space, namely the gradient being parallel [$\bm{k}_{j,+}(\theta_{\bm{r}})$] and antiparallel [$\bm{k}_{j,-}(\theta_{\bm{r}})$] to the observation direction, which yield a significant contribution to both the propagator $G_0(\bm{r},\bm{0};E)$ and the counter-propagator $G_0(\bm{0},\bm{r};E)$ [see Fig.~\ref{fig:scattering} (a)]. In the normal-metal case, only $\bm{k}_{j,+}(\theta_{\bm{r}})$ contributes to the propagator and only $\bm{k}_{j,-}(\theta_{\bm{r}})$ contributes to the counter-propagator. As we discuss below in the analysis of the LDOS, this duality of critical points increases the number and richness of scattering processes. 

Second, the critical points in a normal metal are strictly real and they sit on the Fermi contour. However, in a superconductor, it follows from the in-gap constraint on the propagator's energy (i.e. $E< \Delta \Rightarrow \omega^2 >0$) and Eq.~\eqref{eq:crit_sol_energy} that the critical points $\bm{k}_{j,\pm}(\theta_{\bm{r}})$ are complex numbers. One can observe in Eq.~\eqref{eq:approx} that the real part of the critical points yields the oscillatory behavior of the LDOS, whereas the imaginary part will lead to an exponential decay. Thus, we can define the oscillatory and decay characteristic lengths of the propagator, specifically,
\begin{subequations}
    \begin{align}
        \lambda_{j,\pm}(\theta_{\bm{r}}) &= \frac{1}{\operatorname{Re}[\bm{k}_{j,\pm}(\theta_{\bm{r}})] \cdot \hat{\bm{r}}},\\
        \xi_j(\theta_{\bm{r}}) &= \frac{1}{\operatorname{Im}[\bm{k}_j(\theta_{\bm{r}})] \cdot \hat{\bm{r}}}.   
    \end{align}
    \end{subequations}
The former is reminiscent of the Friedel oscillations in a normal metal, while the latter is the natural consequence of evaluating the bare propagator at sub-gap energies. We note that the critical points of the counter-propagator $G_0(\bm{0},\bm{r};E)$ are the complex-conjugate of the critical points of the propagator $G_0(\bm{r},\bm{0};E)$. Further, we note that owing to the even parity of the energy dispersion we can relate the real and imaginary parts of same-pocket critical points, namely,   $\operatorname{Re}[\bm{k}_{j,+}(\theta_{\bm{r}})] = -\operatorname{Re}[\bm{k}_{j,-}(\theta_{\bm{r}})]$ and $\operatorname{Im}[\bm{k}_{j,+}(\theta_{\bm{r}})] = \operatorname{Im}[\bm{k}_{j,-}(\theta_{\bm{r}})]$. Therefore, we conclude that each pocket contributes to the propagator two terms with the same decay length.

The approximate expression for the bare propagator reads
\begin{align}
\begin{split}
&\hat{G}_0(\bm{r}, \bm{0};E) \sim \frac{1}{\omega\sqrt{r}} \sum_{j,\; \epsilon = \pm}\Gamma_{j,\epsilon}(\theta_{\bm{r}})
\; \cdot\\
&\; \; \cdot e^{-\frac{r}{\xi_j(\theta_{\bm{r}})} + i[\frac{r}{\lambda_{j,\epsilon}(\theta_{\bm{r}})} - \epsilon \frac{\pi}{4}]}
\begin{pmatrix}
E + \epsilon i \omega & \Delta \\
\Delta & E - \epsilon i \omega
\end{pmatrix},
\end{split}\label{eq:prop_approx}
\\[2ex]
\begin{split}
&\text{where} \; 
\Gamma_{j,\epsilon}(\theta_{\bm{r}}) =\frac{1}{|\bm{\nabla}\varepsilon_{\bm{k}_{j,\epsilon}(\theta_{\bm{r}})}|\sqrt{\kappa_{\bm{k}_{j,\epsilon}(\theta_{\bm{r}})}}}.\label{eq:gamma}
\end{split}
\end{align}

In these expressions $|\bm{\nabla}\varepsilon_{\bm{k}_{j, \epsilon}(\theta_{\bm{r}})}|$ and $\kappa_{\bm{k}_{j,\epsilon}(\theta_{\bm{r}})}$ denote the norm of $\bm{\nabla}\varepsilon_{\bm{k}} \equiv \left( \partial_{k_x} \varepsilon_{\bm{k}}, \partial_{k_y} \varepsilon_{\bm{k}}\right)$  and the curvature of $\varepsilon_{\bm{k}} = 0$ evaluated at $\bm{k}_{j, \epsilon}(\theta_{\bm{r}})$ -they are therefore complex numbers. The summation in Eq.~\eqref{eq:prop_approx} accounts for multiple critical points.

We emphasize that the observation direction determines the set of critical points $\bm{k}_{j,\epsilon}(\theta_{\bm{r}})$ through the gradient equation \eqref{eq:crit_sol_gradient}. Therefore, the anisotropy of the LDOS at the YSR-state energy is encoded in the exponential decay and in the oscillation period, as well as in an overall prefactor which depends inversely on the curvature and the norm gradient of the energy dispersion. Under the assumption of a non-vanishing curvature, we obtain that the power-law decay of the LDOS of the YSR state is isotropic and it goes as $1/r$. We conclude that in generic situations solely the substrate dimensionality determines the power-law, while  exceptional behavior can occur if the observation direction is perpendicular to a strictly linear segment of the Fermi surface (then the segment forms a continuum of critical points, with vanishing curvature for each), or if the observation direction has critical points lying on zero-curvature points of the Fermi surface (arguably this leads to a slower algebraic decay \cite{lounis2011theory_real_imaging}).

We note that owing to the even parity of the energy dispersion $\Gamma_{j,+}(\theta_{\bm{r}}) = \Gamma_{j,-}^{*}(\theta_{\bm{r}})$, thus both critical points ($\pm$) belonging to a given pocket $j$ contribute a term with equal amplitude and exponential decay to the propagator.

We remark that the LDOS inherits its anisotropic features from the Fermi contour, therefore, our approximate expression for the bare propagator \eqref{eq:approx} together with the knowledge of an arbitrary energy dispersion $\varepsilon_{\bm{k}}$ allows to predict the orientation and shape of the YSR state. We leave this discussion to Section \ref{subsec:small_gap} where we provide a physical interpretation of the real and imaginary parts of the critical points in terms of the energy dispersion in the limit of a small superconducting gap and we provide a few examples. Next, we continue discussing the scattering processes involved in the LDOS.

\subsection{Underlying scattering mechanisms}
\label{subsec:scatt}
In order to interpret the significance of the critical points $\bm{k}_{j,\pm} (\theta_{\bm{r}})$ it is insightful to write explicitly the product $ \delta \hat{G}(\bm{r},\bm{r}; E) \sim \hat{G}_0(\bm{r}, \bm{0};E) \hat{T}(E) \hat{G}_0(\bm{0}, \bm{r};E)$ up to linear order in the impurity potential. For concreteness we present the electron-electron component which corresponds to the LDOS measured at positive bias; the hole-hole entry is analogous up to a phase factor. The full expression can be found at the end of Appendix \ref{app:calc}. The relevant term contributed by Fermi pockets $j$ and $j'$ is
\begin{equation}
\begin{split}
\label{eq:delta_G}
    \delta G_{\mathrm{ee}}^{j,j'} \sim &\frac{1}{r} \;\sum_{\epsilon,\epsilon'=\pm} \Gamma_{j,\epsilon}(\theta_{\bm{r}}) \Gamma_{j',\epsilon}(\theta_{\bm{r}}) \\
    &\cdot e^{-\frac{r}{\xi_{j,j'}(\theta_{\bm{r}})} + i\frac{r}{\lambda_{j,j'}^{\epsilon, \epsilon'}(\theta_{\bm{r}})}}  G_{0_{\mathrm{e},\alpha}}^{\epsilon} G_{0_{\alpha, \mathrm{e}}}^{\epsilon'},
\end{split}
\end{equation}
where
\begin{subequations}
\begin{align}
    \xi_{j,j'}(\theta_{\bm{r}}) = \left(\frac{1}{\xi_j(\theta_{\bm{r}})} + \frac{1}{\xi_{j'}(\theta_{\bm{r}})}\right)^{-1}, \label{eq:xi_harmonic}\\
    \lambda_{j,j'}^{\epsilon, \epsilon'}(\theta_{\bm{r}}) = \left(\frac{1}{\lambda_{j,\epsilon}(\theta_{\bm{r}})} - \frac{1}{\lambda_{j',\epsilon'}(\theta_{\bm{r}})}\right)^{-1}. \label{eq:lambda_harmonic}
\end{align}
\end{subequations}
The summation in $\alpha$ runs over particle-hole space. The products of the matrix entries read
\begin{subequations}
\begin{align}
      G^{\epsilon}_{0_{\mathrm{e,e}}}G^{\overline{\epsilon}}_{0_{\mathrm{e,e}}} &= (E+\epsilon i \omega)^2, \label{eq:process_ee_2}\\
    G^{\epsilon}_{0_{\mathrm{e,h}}}G^{\epsilon'}_{0_{\mathrm{h,e}}} &= \Delta^2,\label{eq:process_eh}\\
    G^{\epsilon}_{0_{\mathrm{e,e}}}G^{\epsilon}_{0_{\mathrm{e,e}}} &= \Delta^2\label{eq:process_ee_1}.
\end{align}
\end{subequations}

A term shown in Eq.~\eqref{eq:delta_G} represents one of the possible electron-electron scattering processes up to linear order in the impurity potential. Let us start by considering the case of a single-pocket Fermi contour, and thereby drop the summation in $j,j'$.

\textit{Case of a single-pocket Fermi contour}, $j=1$.
As we discussed in the context of the bare propagator, the pair of critical points belonging to the same pocket yields states with the same decay length $\xi_1$. Therefore, in this case all scattering processes have the same decay length $\xi_{1,1}$. Further, scattering processes which reverse the momentum of the excitation, i.e. from $\bm{k}_{1,\epsilon} (\theta_{\bm{r}})$ to $\bm{k}_{1,\overline{\epsilon}} (\theta_{\bm{r}})$ exhibit an oscillatory character controlled by $\lambda_{1,1}^{\epsilon,\overline{\epsilon}}(\theta_{\bm{r}})$. Within this class, we can distinguish a conventional scattering process [Eq.~\eqref{eq:process_ee_2}, Fig.~\ref{fig:scattering} (c) top] and a condensate-mediated scattering process [Eq.~\eqref{eq:process_eh}]. By taking the $\Delta \rightarrow 0$ limit while keeping the energy of the propagator finite it can be observed that the former is reminiscent of the normal-metal scattering while the latter  arises due to the superconducting nature of the substrate. On the other hand, scattering processes which conserve the momentum of the excitation, i.e. from $\bm{k}_{1,\epsilon} (\theta_{\bm{r}})$ to $\bm{k}_{1,\epsilon} (\theta_{\bm{r}})$, do not exhibit an oscillatory character $[\lambda_{1,1}^{\epsilon,\epsilon}(\theta_{\bm{r}})^{-1} = 0]$. All processes belonging to this class are mediated by the superconducting condensate and therefore their amplitude scales with $\Delta^2$ [Eq.~\eqref{eq:process_eh}, Fig.~\ref{fig:scattering} (c) bottom, and Eq.~\eqref{eq:process_ee_1}]. This is consistent with our previous discussion on the critical points, where we pointed out that in the normal-metal scenario only $\bm{k}_+(\theta_{\bm{r}})$ and $\bm{k}_-(\theta_{\bm{r}})$ contribute to the propagator and counter-propagator respectively. Therefore we conclude that momentum-conserving scattering processes are a distinctive feature of the superconducting medium.

\textit{Case of a multi-pocket Fermi contour}, $j>1$.
If there is more than one pocket in the Fermi contour, the discussion of the previous paragraph applies to all \textit{intra}pocket processes. Each pocket contributes to the propagator eight terms which decay with $\xi_{j,j}(\theta_{\bm{r}})$. Now there also exist \textit{inter}pocket scattering processes which decay with $\xi_{j,j'}(\theta_{\bm{r}})$ and oscillate with $\lambda_{j,j'}^{\epsilon, \epsilon'}(\theta_{\bm{r}})$ [see Fig.~\ref{fig:scattering} (b)]. Note that since the LDOS decay length is the harmonic mean of the propagator decay lengths from the existing pockets [Eq.~\eqref{eq:xi_harmonic}], the largest $\xi_{j,j'}(\theta_{\bm{r}})$ always belongs to an intrapocket process, i.e. $j=j'$. Furthermore, in general all interpocket processes have an oscillating character even if $\epsilon = \epsilon'$. As predicted for the normal metal, here the existence of several characteristic frequencies also gives rise to a beating pattern. However, the fact that the largest decay length corresponds to an intrapocket process which has one characteristic frequency only implies that in the very large $|\bm{r}|$ limit the beating pattern will be suppressed. Finally, we remark that the classification of the scattering processes into normal-metal-like and condensate-mediated discussed previously applies to interpocket processes as well. 

\subsection{Small-gap limit}
\label{subsec:small_gap}
To elucidate the meaning of complex critical momenta it is useful to reconcile the normal-metal solution with the superconductor counterpart. As we discussed at the beginning of Sec.~\ref{sec:results}, if $\Delta$ is strictly zero, the critical points are real and sit on the Fermi surface. In Appendix \ref{app:small_gap} we show that in the limit $\Delta \rightarrow 0$, 
\begin{align}
\label{eq:realPart}
 \operatorname{Re}[\bm{k}_{j,\pm}(\theta_{\bm{r}})]&\sim \pm \widetilde{\bm{k}}_{j}(\theta_{\bm{r}}),\\
 \label{eq:imagPart}
   \frac{1}{\operatorname{Im}[\bm{k}_{j}(\theta_{\bm{r}})] \cdot \hat{\bm{r}}} &\equiv \xi_j(\theta_{\bm{r}}) \sim \frac{|\bm{\nabla}\varepsilon_{\widetilde{\bm{k}}_{j}}(\theta_{\bm{r}})|}{\omega},
\end{align}
where $\widetilde{\bm{k}}_{j}(\theta_{\bm{r}})\in \mathbb{R}^2$ is the normal-metal critical point, i.e. a point lying on the Fermi contour where the gradient of the energy dispersion lies parallel to $\hat{\bm{r}}$.

The exponential decay of each pocket is hence given by its anisotropic Fermi velocity. This result provides a transparent generalization of previous analytical studies which assumed an isotropic energy dispersion and found that the LDOS decays with the superconducting coherence length, $\xi_{\mathrm{iso}} \sim \frac{\hbar v_{F}}{\Delta}$ \cite{rusinov, menard:coherent}.

The second source of anisotropy in the propagator is the prefactor $\Gamma_{j,\epsilon}(\theta_{\bm{r}})$ which itself depends on two quantities [see Eq.~\eqref{eq:gamma}]: it goes inversely with the norm of the gradient of the energy dispersion, and inversely with the curvature, both evaluated at the critical point. The inverse curvature causes a phenomenon discussed in the context of charge impurities in three-dimensional metals \cite{heinze2000realimaging, lounis2011theory_real_imaging}: \textit{quasiparticle focusing}. Namely, the inverse curvature is highest on the flattest parts of the Fermi surface, and the prefactor $\Gamma$ will be enhanced for observation directions perpendicular to such segments. The saddle point approach makes this connection explicit: if the observation direction is perpendicular to such a flatter segment, and hence it is aligned with the energy gradient there, the critical point will indeed be on the segment [see Eq.~\eqref{eq:crit_sol_gradient}] and its inverse curvature will be high. The quasiparticle focusing in our theory for superconductors hence justifies why previous experimental works show an enhancement of the LDOS along directions perpendicular to the flattest segments of the Fermi surface \cite{menard:coherent, wiesendanger:focusing, thupakula2021coherent}. 

The norm of the gradient of the energy dispersion plays a crucial role in the anisotropy of the YSR LDOS. Firstly, through the decay length which is enhanced where the gradient is the highest, as we discussed in the
beginning of this subsection. Previous studies of superconductors failed to point out this dependence, which constitutes a fundamental difference with respect to the normal-metal scenario where the impurity response lacks any exponential decay length. Secondly, as a quantity entering \textit{inversely} in the prefactor $\Gamma_{j,\epsilon}(\theta_{\bm{r}})$, which stems from the reduced dimensionality of the substrate (we find the same prefactor in a two-dimensional normal metal, see App.~\ref{app:nm}). Hence the norm of the gradient reduces the prefactor $\Gamma_{j,\epsilon}(\theta_{\bm{r}})$ for observation directions for which it enhances the decay length, and naively one would expect a competition. Nevertheless, we observed in all studied examples that overall the prefactor $\Gamma_{j,\epsilon}(\theta_{\bm{r}})$ and the characteristic length $\xi_{j}(\theta_{\bm{r}})$ grow and shrink in phase as the observation direction $\theta_{\bm{r}}$ varies. This behavior is possible because the reduction of $\Gamma$ due to the inverse norm gradient can be more than compensated by the inverse curvature. In Appendix \ref{app:pref} we provide a scaling argument to justify that indeed overall $\Gamma_{j,\epsilon}(\theta_{\bm{r}})$ varies as the inverse curvature (e.g., it is highest on the flattest segments of the Fermi surface).

\subsubsection{Application to a single-pocket model}
\label{subsec:single}

\begin{figure}
    \centering
    \includegraphics[width=0.99\columnwidth]{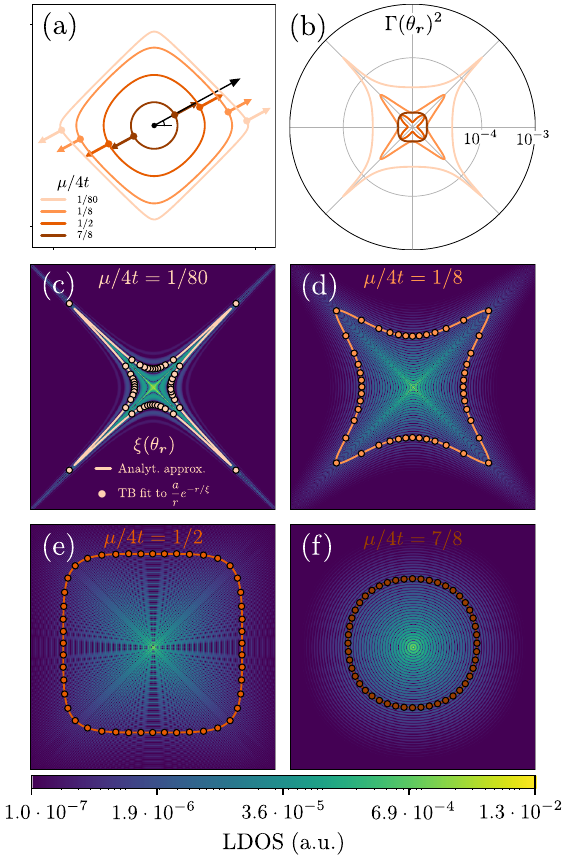}
    \caption{(a) Fermi contours of the energy dispersion in Eq. \eqref{eq:ek_tb_sqrl} at several doping values. The long, black arrow indicates an arbitrary observation direction $\theta_{\bm{r}}$, and the small, colored arrows the normalized gradient of the energy dispersion at the corresponding critical points. Note that for a perfectly circular Fermi contour the critical point would sit at $\theta_{\bm{r}}$ exactly. (b) Polar plot of the LDOS prefactor in log-scale for the energy dispersions represented in (a). (c-f) Electron part of the LDOS at the YSR-state energy calculated numerically from the energy dispersions in (a). Field of view is 401 by 401 lattice sites around the impurity. Color bar in arbitrary units, log-scale. The color curve is a polar plot of the decay length. The solid line indicates the analytical approximation and the circular markers are extracted from fitting cuts of the numerical LDOS. We note that the scale of the polar curves is different from the scale of the underlying color maps; the circumscribing circle of the color curve in (c) corresponds to 65 lattice sites. Numerical parameters: $t= 200$ meV, $\Delta = 5$ meV, $J = 285$ meV.}
    \label{fig:single_pocket}
\end{figure}

To illustrate the relationship between the Fermi contour and the anisotropy of the YSR states, let us consider a nearest-neighbors tight-binding energy dispersion on a square lattice,
\begin{equation}
    \label{eq:ek_tb_sqrl}
    \varepsilon_{\bm{k}} = \mu -2t(\cos k_x + \cos k_y),
\end{equation}
where $\mu$ is the chemical potential and $t$ is the hopping amplitude. As we tune the chemical potential away from the mid-band point the Fermi contours become more isotropic [see Fig.~\ref{fig:single_pocket} (a)], and so does the LDOS at the YSR-state energy [see color maps in Fig.~\ref{fig:single_pocket} (c-f)].

For a given doping, the LDOS prefactor is most prominent along directions perpendicular to the flattest sections of the Fermi contour, namely $\theta_{\bm{r}} = \pm \frac{\pi}{4}$ [Fig.~\ref{fig:single_pocket} (b)]. Nevertheless,  we recall that the prefactor does not only depend on the curvature of the Fermi contour, but also on the inverse of the angle-dependent Fermi velocity. Compare, for instance, the pale-orange ($\mu/t = 1/80$) and brown ($\mu/t = 7/8$) curves in Fig.~\ref{fig:single_pocket} (a) and (b) at $\theta_{\bm{r}} = 0$. Although for that direction the curvature of the $\mu/t = 1/80$ contour is larger, the Fermi velocity is substantially smaller, leading to a larger prefactor.

On the other hand, the exponential decay length [color line in Fig.~\ref{fig:single_pocket} (c-f)] is wholly governed by the angle-dependent Fermi velocity and as we argue in Appendix \ref{app:pref}, it is in phase with the prefactor.

Finally, we remark the excellent agreement between the decay length calculated with the small-gap analytical approximation and the decay length extracted from the tight-binding calculation [solid color line and markers in Fig.~\ref{fig:single_pocket} (c-f) respectively]. To obtain the former we find the critical points lying on the Fermi surface and fulfilling $\bm{\nabla}\varepsilon_{\bm{k}} \parallel \hat{\bm{r}}$, and subsequently we evaluate expression \eqref{eq:imagPart}. To obtain the latter, we compute the LDOS at the YSR-state energy for the energy dispersion \eqref{eq:ek_tb_sqrl} numerically \cite{menard:coherent} and we extract the decay length from fitting radial cuts to the envelop function of the LDOS, namely $\frac{a}{r}e^{-r/\xi_{\mathrm{LDOS}}}$. In the next subsection we perform the same analysis of a realistic tight-binding model, thereby showing all the power of the analytical approximation.

\subsubsection{Application to a multi-pocket model}
\label{subsec:multi}
\begin{figure}
    \centering
    \includegraphics[width=0.85\columnwidth]{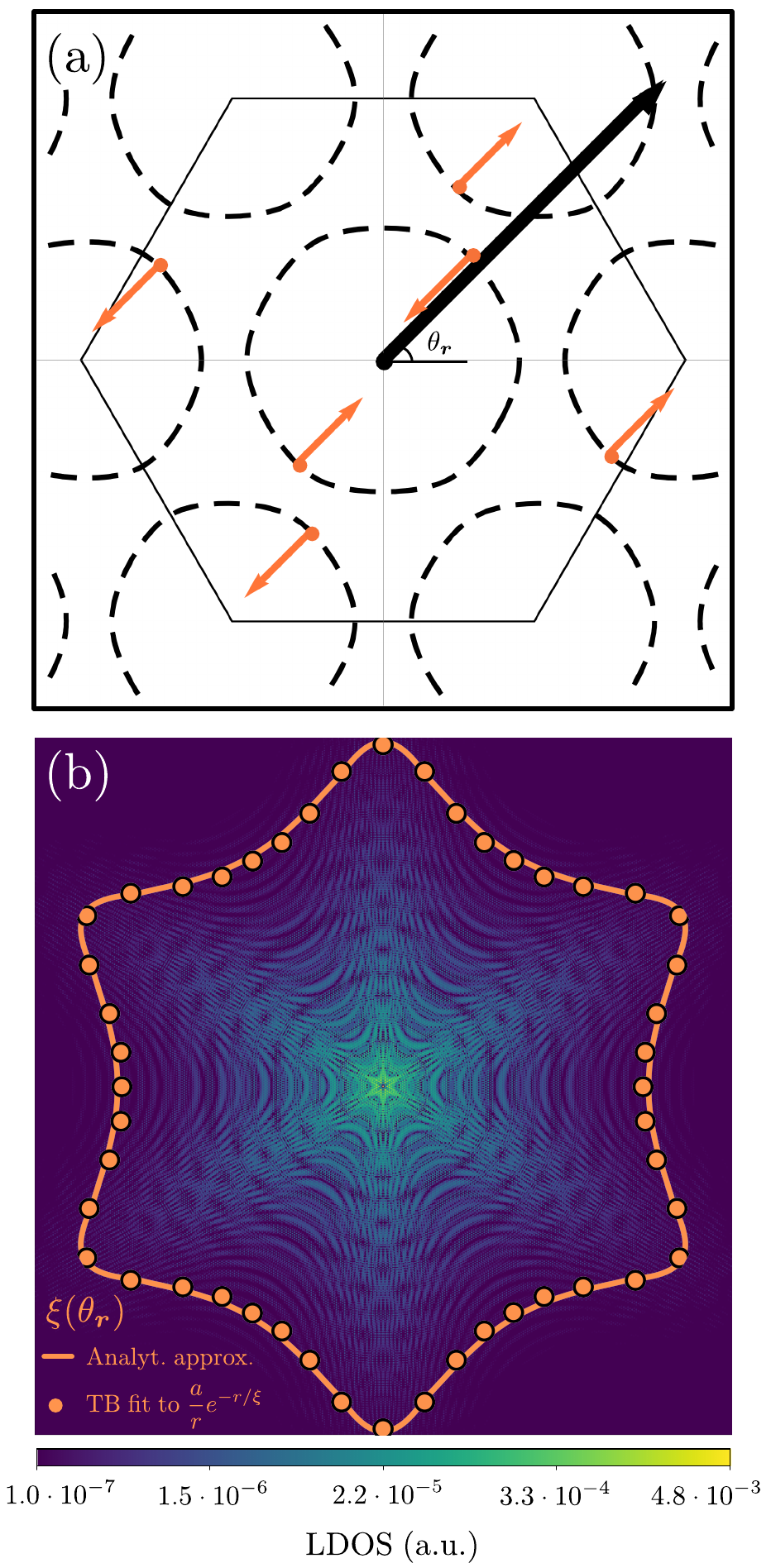}
    \caption{(a) Fermi surface for fifth-nearest neighbors tight-binding model on a triangular lattice describing NbSe\textsubscript{2}, parameters from band 2 in \cite{rahn2012arpes}. Black hexagon marks the first Brillouin zone. The long, black arrow indicates an arbitrary observation direction $\theta_{\bm{r}}$, and the small, orange arrows mark the gradient of the energy dispersion evaluated at the critical points on the Fermi surface for the choice of $\theta_{\hat{\bm{r}}}$. (b) Same as panels (c-f) in Fig.~\ref{fig:single_pocket} for the NbSe\textsubscript{2} energy dispersion, with a field of view of 500 by 500 lattice sties. The circumscribing circle of the orange curve corresponds to 59 lattice sites.}
    \label{fig:multi_pocket}
\end{figure}

Next, we apply the approximation presented above to a fifth-nearest neighbors tight-binding energy dispersion on a triangular lattice, which is known to faithfully describe some monolayer transition metal dichalcogenides, such as NbSe\textsubscript{2} \cite{rahn2012arpes, menard:coherent}. The energy dispersion evaluated at the hopping parameters extracted from best-fitted NbSe\textsubscript{2} yields a disconnected Fermi surface which has three non-equivalent Fermi pockets, specifically at $\Gamma$, $K$ and $K'$ points. Therefore, for a given observation direction $\theta_{\bm{r}}$ we have three pairs of critical points [see Fig.~\ref{fig:multi_pocket} (a)].

As predicted by our theory, the LDOS at the YSR-state energy is enhanced along directions perpendicular to flatter sections of the Fermi contours [Fig.~\ref{fig:multi_pocket} (b)]. Remarkably the analytical approximation for the exponential decay length and the numerical fits are also in excellent agreement in this case, despite the complexity of the Fermi surface. As we discussed in Sec.~\ref{subsec:scatt}, an $N_j = 3$ Fermi contour, yields six different decay lengths. The color line in Fig.~\ref{fig:multi_pocket} (b) represents the largest $\xi_{j,j'}$, nevertheless, we note that for the present energy dispersion, the difference between the various decay lengths is of a few lattice sites only, and therefore negligible. This example showcases the ability of this method to predict the shape and orientation of a YSR state on an arbitrary substrate.

\subsection{Application to extended $s$-wave pairing}
\label{subsec:extended}
Up to this point we restricted our considerations to conventional $s$-wave superconductors. Nevertheless, the formalism developed in the previous sections allows to treat more involved situations where the superconducting gap function is momentum dependent. In order to conserve the structure of the Green functions \eqref{eq:bare_prop_1}, we will stick to singlet pairing and simply incorporate a $\bm{k}$ in $\Delta$, but we emphasize that in principle the technique could be employed in arbitrarily gapped superconductors. Note that $\Delta_{\bm{k}}$ must be an even function as required by the fermionic anticommutation rules. It is useful to introduce the BdG energy dispersion, $E_{\bm{k}} = \sqrt{\varepsilon_{\bm{k}}^2 + \Delta_{\bm{k}}^2}$ (not to be confused with $E$, the energy of the propagator), which allows to express the critical-point conditions in a compact form:
\begin{subequations}
\label{eq:crit_sol_dK}
\begin{align}
    E_{\bm{k}'_{j, \pm} (\theta_{\bm{r}})} &= \pm E,\\
    \bm{\nabla}E_{\bm{k}'_{j,\pm}(\theta_{\bm{r}})} &= \pm{|\bm{\nabla}E_{\bm{k}'_{j,\pm}}|} \hat{\bm{r}(\theta_{\bm{r}})}.
\end{align}
\end{subequations}
Naturally, Eqs.~\eqref{eq:crit_sol} reduce to Eqs.~\eqref{eq:crit_sol_dK} if $\Delta$ is independent of $\bm{k}$. The setting discussed earlier can be formally understood as a particular case of the present situation; however, formulating the solution for $\bm{k}$-independent pairing in terms of the normal electron energy dispersion and the actual Fermi surface was more illuminating and permitted a clearer physical interpretation.

The structure of the bare propagator is analogous to the solution discussed in the preceding sections. The power-law goes as $1/\sqrt{r}$ as dictated by the dimensionality of the substrate, and the anisotropy is encoded in the argument of the exponential function and in the prefactor $\Gamma'_{j,\epsilon}(\theta_{\bm{r}}) =\frac{1}{|\bm{\nabla}E_{\bm{k}'_{j,\epsilon}(\theta_{\bm{r}})}|\sqrt{\kappa_{\bm{k}'_{j,\epsilon}(\theta_{\bm{r}})}}}$. Now the curvature and the norm of the gradient refer to the BdG dispersion $E_{\bm{k}}$.

To further understand the effect of a nodeless, anisotropic superconducting gap on the spatial structure of the YSR state, it is insightful to treat the $\bm{k}$ dependent part as a perturbation of a constant background,
\begin{equation}
    \Delta_{\bm{k}} = \Delta + \Delta' f_{\Delta} (\bm{k}),
\end{equation}
with $\Delta' \ll \Delta$ and $f_{\Delta} (\bm{k})$ an even function of $\bm{k}$, and further, to consider the small-gap limit discussed in Section \ref{subsec:small_gap}. We find that the exponential decay length is corrected as follows
\begin{equation}
 \label{eq:imagPart_dK}
  \xi'_{j,\epsilon}(\theta_{\bm{r}}) \sim \frac{|\bm{\nabla}\varepsilon_{\widetilde{\bm{k}}_{j}}(\theta_{\bm{r}})|}{\omega} \left(1 + \frac{\Delta \Delta'}{\omega^2}f_\Delta[\widetilde{\bm{k}}_{j,\epsilon}(\theta_{\bm{r}})]\right)^{-1},
\end{equation}
where $\widetilde{\bm{k}}_{j,\epsilon} (\theta_{\bm{r}})\in \mathbb{R}^2$ is the critical point in the normal metal limit. 

\begin{figure*}
    \centering
    \includegraphics[width=1.0\textwidth]{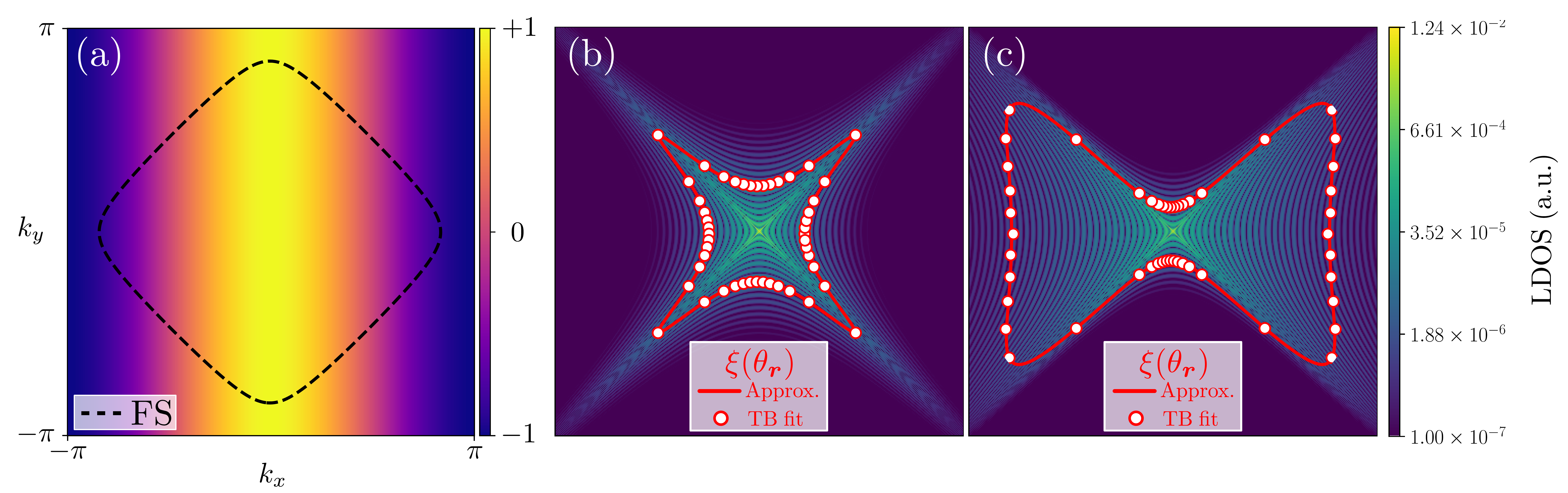}
    \caption{(a)  The color map  represents $f_{\Delta}(\bm{k})$ on the FBZ. The dashed line represents the Fermi contour of the tight-binding energy dispersion \eqref{eq:ek_tb_sqrl} with $\mu/t = 0.25$. (b) and (c) Same as panels (c-f) in Fig.~\ref{fig:single_pocket} for the tight-binding model discussed in the present section. In panel (b) $\Delta' = 0$ while in panel (c) $\Delta'/\Delta = 0.8$. The field of view of the LDOS plots is 401 x 401 sites while the largest $\xi_{\mathrm{LDOS}}$ in the red curves is $\sim$ 83 sites. Numerical parameters: $t = 200$ meV, $\Delta = 5$ meV, $J = 285$ meV.}
    \label{fig:extended_swave}
\end{figure*}

In order to exemplify this result we benchmark the analytical approximation for the decay length against a tight-binding calculation of the LDOS at the YSR-state energy. We choose the energy dispersion introduced in Eq. \eqref{eq:ek_tb_sqrl} describing a nearest-neighbours tight-binding model on a square lattice. Further, we take a nearest-neighbours superconducting coupling such that
\begin{equation}
    \label{eq:TB_delta}
    f_{\Delta}(\bm{k}) = \cos k_x + \alpha \cos k_y,
\end{equation}
where $\alpha \in [0,1]$ is a parameter to control the anisotropy. In the limit $\alpha =1$ this pairing function is known under the name of extended $s$-wave or unconventional $s_{\pm}$-wave and it has been proposed as a candidate to describe iron-based superconductors \cite{MAZIN2009614, mashkoori2019}. However, we choose  $\alpha = 0$ in our calculations to maximize the variation of the gap along the Fermi contour, thereby enhancing the effect of the pairing function anisotropy on the LDOS of the YSR state.

Results are presented in Figure \ref{fig:extended_swave}. In the absence of gap-anisotropy [$\Delta'=0$, panel (b)] the LDOS naturally exhibits the four-fold rotational symmetry of the underlying lattice model. As discussed in Sect. \ref{subsec:single}, the decay length is the most prominent along the $\theta_{\bm{r}} = \pm \frac{\pi}{4}$ directions for which the gradient of the energy dispersion is the largest. When we switch on an anisotropic texture on the superconducting gap the symmetry of the LDOS is reduced accordingly [see panel (c) in Fig.~\ref{fig:extended_swave}]. We recall that within the working approximation, the critical points sit on the Fermi contour, therefore, if we set the observation direction along $\theta_{\bm{r}}=0$ for instance, we have that $f_{\Delta}[\widetilde{\bm{k}}(0)] < 0$ [see panel (a) in Fig.~\ref{fig:extended_swave}]. This leads to an enhancement of the decay length as predicted in Eq.~\eqref{eq:imagPart_dK}. We note that the approximation captures very well the intricacies of the spatial structure of the LDOS despite the seemingly large value of $\Delta'/\Delta$ employed in the numerical calculations. 

\section{Conclusions}
\label{sec:concl}
In this work we provide a precise explanation of the role of the Fermi surface in the spatial anisotropy of YSR states in two-dimensional $s$-wave superconductors. To summarize, the anisotropy of the LDOS is encoded in an overall prefactor, and in the exponential decay length and oscillations. The prefactor also arises in the charge-density response in normal metals, and it depends inversely on the angle-dependent Fermi velocity and on the curvature of the Fermi contour, meaning that YSR states show prominent features along directions perpendicular to flatter sections of the Fermi contours. The decay length is proportional to the angle-dependent Fermi velocity constituting an elegant generalization of the superconducting coherence length which governs the exponential decay of YSR states on isotropic substrates. Through a simple scaling argument we show that the prefactor and the decay length are always in phase, therefore the knowledge of the energy dispersion allows to predict the shape and orientation of YSR states, even for STM measurements whose field of view is too small to encompass the exponential decay. Understanding how the Fermi surface shapes the spatial structure of YSR states eases the path towards the optimal design of collective impurity states on superconductors.

We emphasize that contrary to previous works \cite{ortuzar2021yushibarusinov} we do not make any approximations regarding the Fermi surface, but instead, we apply our analytical expression to arbitrarily complex energy dispersions.  Aside from reproducing the symmetry of the YSR states reported in STM experiments on NbSe\textsubscript{2}, we achieve an accurate \textit{quantitative} comparison with a tight-binding calculation using a realistic energy dispersion for NbSe\textsubscript{2} which showcases the power of our analytical approximation. Further, we find that under the assumption of a Fermi surface with strictly positive curvature, the power-law decay of the LDOS at the YSR-state energy goes as $1/r$ and depends on the substrate dimensionality only. Earlier works \cite{wiesendanger:focusing, ortuzar2021yushibarusinov} had suggested that quasiparticle focusing could lead to a slower decay. It remains an open question to investigate the next leading terms in the saddle-point expansion, which will arguably have a slower algebraic decay. These terms will become relevant for observation directions for which the critical points lie on straight segments of the Fermi surface or with nearly vasnishing curvature. Incidentally, the method presented in Ref.~\cite{ortuzar2021yushibarusinov} becomes more accurate in this limit, therefore, we conlude that the two theoretical descriptions complement each other.

Moreover, we find that the most likely scattering processes involve excitations whose momenta lies on the Fermi contour where the gradient of the normal energy dispersion is parallel and anti-parallel to the observation direction. This implies that each pocket of the Fermi surface contributes twice as many meaningful scattering momenta than  in the normal-metal scenario. The emerging scattering processes are indeed mediated by the condensate and constitute a distinctive feature of the superconducting nature of the substrate. Unfortunately, all the contributions to the bare propagator stemming from the same pocket decay similarly, therefore it does not seem plausible to arbitrarily enhance the Andreev-like processes; nevertheless, our analysis via the saddle-point technique deepens the current understanding of the underlying scattering mechanisms in the YSR problem.

Finally, we demonstrate that the analytical approximation also offers quantitatively correct results in superconductors with a momentum-dependent pairing function. This opens the door to study more complex situations such as $p$-wave superconductors or multi-band substrates. 

\appendix
\section{Charge impurity in a 2D normal metal}
\label{app:nm}

In this Appendix we extend the result derived in \cite{lounis2011theory_real_imaging} to a two-dimensional substrate. The energy-resolved change in charge density due to a scalar impurity embedded in the substrate is given by
\begin{equation}
    \label{eq:app-charge}
    \delta n(\bm{r}; E) \sim \operatorname{Im}[G_0(\bm{r},\bm{0};E) T(E) G_0(\bm{0},\bm{r};E)],
\end{equation}
where contrary to the superconducting case discussed in Sec.~\ref{sec:model} the bare propagator $G_0$ and the transfer ``matrix'' $T$ are scalar. Analogously, under the assumption of a point-like impurity, all the spatial information is encoded in the bare propagator, therefore we will perform a saddle-point approximation to calculate the integral
\begin{equation}
G_0(\bm{r},\bm{0};E) = \int \frac{d\bm{k}}{(2\pi)^2}\frac{e^{i\bm{k}\cdot\bm{r}}}{E+i0^+ - \varepsilon_{\bm{k}}},    
\end{equation}
and its counterpart $G_0(\bm{0},\bm{r};E)$. We introduce an auxiliary variable $t$ and write
\begin{equation}
    \label{eq:app-G0_t_NM}
    G_0(\bm{r},\bm{0};E) = -i |\bm{r}|\int \frac{d\bm{k}}{(2\pi)^2}\int_0^{\infty} dt e^{i|\bm{r}|\phi(\bm{k},t)},
\end{equation}
where
\begin{equation}
    \label{eq:app-phase}
    \phi(\bm{k},t) = \bm{k}\cdot \hat{\bm{r}} + t(E+i0^+ - \varepsilon_{\bm{k}}).
\end{equation}
The critical points $(\bm{k}_j, t_j)$ giving the largest contribution to the the integral \eqref{eq:app-G0_t_NM} fulfill $(\bm{\nabla}, \partial_t)\phi(\bm{k},t) = 0$, and naturally, they depend on the observation direction $(\theta_{\bm{r}})$. This yields the following conditions
\begin{subequations}
    \begin{align}
    \label{eq:app-energy_eq_NM}
        \varepsilon_{\bm{k}_j(\theta_{\bm{r}})} &= E,\\
    \label{eq:app-grad_eq_NM}
        \bm{\nabla}\varepsilon_{\bm{k}_j(\theta_{\bm{r}})} &= |\bm{\nabla}\varepsilon_{\bm{k}_j(\theta_{\bm{r}})}|\hat{\bm{r}},\\
    \label{eq:app-t_eq_NM}
        t_j(\theta_{\bm{r}}) &= \frac{1}{|\bm{\nabla}\varepsilon_{\bm{k}_j(\theta{\bm{r}})}|}.
    \end{align}
\end{subequations}
Note that Eq.~\eqref{eq:app-energy_eq_NM} follows from taking the limit $\Delta \rightarrow 0$ while keeping $E$ finite in the superconductor critical equation \eqref{eq:crit_sol_energy}. Contrary to the superconductor scenario, the ``anti-parallel'' solution does not contribute to the integral in the normal metal case [compare \eqref{eq:crit_sol_gradient} and \eqref{eq:app-grad_eq_NM}]. Further, we remark that Eq.~\eqref{eq:app-energy_eq_NM} yields an iso-energy contour at the propagator energy, the relevant contour corresponding to $E=0$, i.e., the Fermi contour. Finally, we note that the counter propagator $G_0(\bm{0},\bm{r};E)$ yields an analogous set of equations up to a minus sign in \eqref{eq:app-grad_eq_NM}, thereby spawning the shaded entries in Fig.~\ref{fig:scattering}, panel (b).

From now on, it is understood that critical points depend on the observation direction, therefore we drop $(\theta_{\bm{r}})$ to lighten the notation. Next, we expand the phase \eqref{eq:app-phase} up to second order around the critical points. The zeroth order contribution follows trivially,
\begin{equation}
    G_0(\bm{r},0;E) \sim -i |\bm{r}| e^{i \bm{k}_j \cdot \bm{r}}.
\end{equation}
To second order, the integral in $t$ reads
\begin{equation}
\begin{split}
\label{eq:app-int_t}
    \int_0^{\infty}& d \Delta t_j \exp\left[-i |\bm{r}|\frac{1}{2}\sum_{\alpha}\frac{\partial \varepsilon_{\bm{k}}}{\partial k_\alpha}\Bigr \rvert_{\bm{k}_j} \Delta k_{\alpha_j} \Delta t_j\right]\\
    &\sim \frac{4\pi}{|\bm{r}||\bm{\nabla}\varepsilon_{\bm{k}_j}|} \delta(\Delta k_{\hat{\bm{r}}_j}),
\end{split}
\end{equation}
where $\Delta \cdot_j$ are the integration variables which denote a small interval around the critical point $\cdot_j$, $\delta$ represents the Dirac delta distribution, and $k_{\hat{\bm{r}}_j}$ is the projection of $\bm{k}_j$ along $\hat{\bm{r}}$. To approximate \eqref{eq:app-int_t} we extended the integration bounds from $[0, \infty)$ to $(-\infty, \infty)$ and we used the fact that $\delta (\nabla \varepsilon_{\bm{k}_j} \cdot \Delta \bm{k}_j) = \frac{\delta(\Delta k_{\hat{\bm{r}}_j})}{|\bm{\nabla}\varepsilon_{\bm{k}_j}|} $ which follows from the critical-point equation \eqref{eq:app-grad_eq_NM}.

Finally, by rotating the integration axes so that they lie tangent and normal to an iso-energy contour \eqref{eq:app-energy_eq_NM}, the integral in $\bm{k}$ reads
\begin{equation}
\begin{split}
    \int_{-\infty}^{\infty} & d\Delta k_{\hat{\bm{r}}_j} d\Delta k_{\hat{\bm{r}}_{\perp_j}}\delta(\Delta k_{\hat{\bm{r}}_j}) \exp\Biggl[-i |\bm{r}|\frac{t_j}{2}\sum_{\alpha \beta} \frac{\partial^2 \varepsilon_{\bm{k}}}{\partial k_\alpha \partial k_\beta}\Bigr \rvert_{\bm{k}_j} \cdot \\
    &\cdot \Delta k_{\alpha_j} \Delta  k_{\beta_j}\Biggr]
    \sim \left( \frac{2}{\pi |\bm{r}| |\bm{\nabla}\varepsilon_{\bm{k}_j}|^2t_j \partial^2_{k_{\hat{\bm{r}}_\perp}^2}\varepsilon_{\bm{k}_j}}\right)^{1/2}e^{i \varphi_j},
\end{split}
\end{equation}
where $\varphi_j = -\frac{\pi}{4}\operatorname{sign}(\partial^2_{k_{\hat{\bm{r}}_\perp}^2}\varepsilon_{\bm{k}_j})$ and $\partial^{2}_{k_{\hat{\bm{r}}_\perp}^2}\varepsilon_{\bm{k}_j}$ is the second directional derivative tangent to the iso-energy contour evaluated at the critical point. Putting everything together
\begin{equation}
\label{eq:app-G0_final_NM}
    G_0(\bm{r},\bm{0};E) \sim \frac{1}{\sqrt{r}} \sum_j \Gamma_{j}(\theta_{\bm{r}})  e^{i[\bm{k}_j(\theta_{\bm{r}})\cdot \bm{r} + \varphi_j]},
\end{equation}
where $\Gamma_{j}(\theta_{\bm{r}}) =\frac{1}{|\bm{\nabla}\varepsilon_{\bm{k}_{j}(\theta_{\bm{r}})}|\sqrt{\kappa_{\bm{k}_{j}(\theta_{\bm{r}})}}}$. In this expression $|\bm{\nabla}\varepsilon_{\bm{k}_{j}(\theta_{\bm{r}})}|$ and $\kappa_{\bm{k}_{j}(\theta_{\bm{r}})}$ denote the norm of the gradient of the energy dispersion and the curvature of the Fermi contour evaluated at the critical point $\bm{k}_{j}(\theta_{\bm{r}})$. The summation in Eq.~\eqref{eq:app-G0_final_NM} accounts for multiple critical points stemming from a multi-pocket energy dispersion.

The bare propagator in the normal metal differs from the bare propagator in the superconducting phase in its lack of an exponential decay, and in that only the $\bm{k}$-points where the gradient points \textit{parallel} to the observation direction contribute to the summation in Eq.~\eqref{eq:app-G0_final_NM}. However, we find the same power-law decay -which, as emphasized in the main text, depends on the dimensionality only-, and an analogous anisotropic prefactor. Interestingly, contrary to the three-dimensional case \cite{lounis2011theory_real_imaging}, now the prefactor also depends inversely on the angle-dependent Fermi velocity $|\bm{\nabla}\varepsilon_{\bm{k}_j(\theta_{\bm{r}})}|$.

\section{Saddle point approach for 2D superconductors}
\label{app:calc}
\subsection{General solution}
In this Appendix we detail the calculation of the Fourier transforms of the bare propagator defined in Section \ref{sec:model}. Since critical points in the saddle-point approximation lie somewhere in the complex plane outside the original integration range, the procedure is slightly different to that of App.~\ref{app:nm}. To perform the approximation we introduce a new integration variable $t$ and re-express the bare propagator \eqref{eq:bare_prop_1} as 
\begin{equation}
\label{eq:ap-integral_t_sc}
G_0^{\alpha \beta}(\bm{r}, \bm{0}; E) =  -|\bm{r}|\int \frac{d\bm{k}}{(2\pi^2)}f_{\alpha \beta}(\bm{k}) \int_0^{\infty} dt \; e^{-|\bm{r}|\phi_{\hat{\bm{r}}}(\bm{k},t)},
\end{equation}
where $f_{\alpha \beta}(\bm{k})$ comprises the matrix entries in Eq.~\eqref{eq:bare_prop_1}, and
\begin{equation}
\label{eq:ap-phase_sc}
 \phi_{\hat{\bm{r}}}(\bm{k}, t) = -i \bm{k}\cdot \hat{\bm{r}} + (\varepsilon_{\bm{k}}^2+\omega^2)t,
\end{equation}
with $\theta_{\bm{r}}$ the polar angle determined by $\hat{\bm{r}}$. The critical points $(\bm{k}_j, t_j)$ giving the largest contribution to the integral \eqref{eq:ap-integral_t_sc} satisfy $(\bm{\nabla}, \partial_t)\phi_{\hat{\bm{r}}}(\bm{k},t) = 0$. For a given radial direction $\theta_{\bm{r}}$ this equation yields the set of conditions \eqref{eq:crit_sol} presented in the main text which we rewrite here for completeness along with an extra condition for the $t$ variable:
\begin{subequations}
\label{eq:ap-crit_sol}
\begin{align}
\varepsilon_{\bm{k}_{j, \pm}(\theta_{\bm{r}})} &= \pm i \omega,\\
\bm{\nabla}\varepsilon_{\bm{k}_{j, \pm}(\theta_{\bm{r}})} &=  \pm|\bm{\nabla}\varepsilon_{\bm{k}_{j, \pm}(\theta_{\bm{r}})}| \hat{\bm{r}},\\
t_{j,\pm} &= \frac{1}{2 \omega |\bm{\nabla}\varepsilon_{\bm{k}_{j, \pm}(\theta_{\bm{r}})}|}.
\end{align}
\end{subequations}

Next, we recall the following compact expression to perform the saddle-point approximation on a multivariate complex integral. For details on its derivation we refer the reader to \cite{Bleistein12saddlepoint}. Let $I_n(\lambda)$ be an integral over $n$ complex variables defined as
\begin{equation}
    I_n(\lambda) = \int d\bm{z} f(\bm{z}) e^{-\lambda \phi(z)}, \quad \bm{z} = (z_1, z_2, \dots, z_n).
\end{equation}
In the ``large'' $\lambda$ limit we have 
\begin{equation}
    I_n(\lambda) \sim \left(\frac{2\pi}{\lambda}\right)^{n/2}\frac{f(\bm{z}_{\mathrm{s}})}{\sqrt{\det\{H_n[\phi(\bm{z}_{\mathrm{s}})]\}}} e^{-\lambda \phi(\bm{z}_{\mathrm{s}})},
\end{equation}
where $\bm{z}_{\mathrm{s}}$ is defined through the equation $\bm{\nabla}_{\bm{z}}\phi(\bm{z})\Bigr\rvert_{\bm{z}_{\mathrm{s}}} = 0$ and $H_n[\phi(\bm{z}_{\mathrm{s}})] = \frac{\partial^2\phi(\bm{z})}{\partial z_\alpha \partial z_\beta}\Bigr\rvert_{\bm{z}_{\mathrm{s}}}, \; \alpha, \beta =1,\dots, n$, as long as $\det\{H_k[\phi(\bm{z}_{\mathrm{s}})]\} \neq 0$ for $1\leq k \leq n$. 

In our problem the large parameter is $|\bm{r}|$, namely the distance from the impurity, $n=3$ with $\bm{z}\equiv (k_x, k_y, t)$, and $f(\bm{z})$ and $\phi(\bm{z})$ are defined in Eqs.~\eqref{eq:ap-integral_t_sc} and \eqref{eq:ap-phase_sc}.

The approximate expression for the bare propagator reads
\begin{equation}
\begin{split}
\label{eq:ap-prop_approx_sc}
G^{\alpha\beta}_0(\bm{r}, &\bm{0};E) \sim \frac{1}{\omega\sqrt{r}} \sum_{j,\; \epsilon = \pm} \\
&\Gamma_{j,\epsilon}(\theta_{\bm{r}}) f_{\alpha \beta}[\bm{k}_{j,\epsilon}(\theta_{\bm{r}})] e^{i[\bm{k}_{j,\epsilon}(\theta_{\bm{r}}) - \epsilon \frac{\pi}{4}]},
\end{split}
\end{equation}
where $\Gamma_{j,\epsilon}(\theta_{\bm{r}})$ is the anisotropic prefactor introduced in Sec.~\ref{sec:results}.

The critical conditions for the counter-propagator $G_0^{\alpha \beta}(\bm{0}, \bm{r}; E)$ read as follows,
\begin{subequations}
\label{eq:ap-crit_sol_counter}
\begin{align}
\varepsilon_{\bm{k}_{j, \pm}(\theta_{\bm{r}})} &= \mp i \omega,\\
\bm{\nabla}\varepsilon_{\bm{k}_{j, \pm}(\theta_{\bm{r}})} &=  \pm|\bm{\nabla}\varepsilon_{\bm{k}_{j, \pm}(\theta_{\bm{r}})}| \hat{\bm{r}},\\
t_{j,\pm} &= \frac{1}{2 \omega |\bm{\nabla}\varepsilon_{\bm{k}_{j, \pm}(\theta_{\bm{r}})}|}.
\end{align}
\end{subequations}
By comparing the sets of equations \eqref{eq:ap-crit_sol} and \eqref{eq:ap-crit_sol_counter} it becomes evident that the critical points of the counter-propagator $G_0^{\alpha \beta}(\bm{0}, \bm{r}; E)$ are in fact the complex-conjugate of the critical points of the propagator $G_0^{\alpha \beta}(\bm{r}, \bm{0}; E)$. It follows,
\begin{equation}
\begin{split}
\label{eq:ap-counter_prop_approx_sc}
G^{\alpha\beta}_0(\bm{0}, &\bm{r};E) \sim \frac{1}{\omega\sqrt{r}} \sum_{j,\; \epsilon = \pm} \\
&\Gamma_{j,\epsilon}(\theta_{\bm{r}})^* f_{\alpha \beta}[\bm{k}_{j,\epsilon}(\theta_{\bm{r}})]^* e^{-i[\bm{k}_{j,\epsilon}(\theta_{\bm{r}})^* - \epsilon \frac{\pi}{4}]},
\end{split}
\end{equation}
where we used the fact that $\Gamma_{j,\epsilon}(\theta_{\bm{r}})$ and $f_{\alpha \beta}[\bm{k}_{j,\epsilon}(\theta_{\bm{r}})]$ are real functions evaluated at complex values.

The full expression for the electron-electron component for the LDOS in terms of the parameters introduced in the main text reads
\begin{widetext}
\begin{equation}
\label{eq:app-delta_G}
    \delta G_{\mathrm{ee}} \sim \frac{1}{r} \sum_{j,j'}  e^{-\frac{r}{\xi_{j,j'}(\theta_{\bm{r}})}} \sum_{\epsilon,\epsilon'} \Gamma_{j,\epsilon}(\theta_{\bm{r}}) \Gamma_{j',\epsilon}(\theta_{\bm{r}}) \exp \left\{i\left[\frac{r}{\lambda_{j,j'}^{\epsilon, \epsilon'}(\theta_{\bm{r}})} - (\epsilon - \epsilon')\frac{\pi}{4}\right]\right\} \sum_{\alpha} G_{0_{\mathrm{e},\alpha}}^{\epsilon} G_{0_{\alpha, \mathrm{e}}}^{\epsilon'}.
\end{equation}
\end{widetext}
\subsection{Small-gap approximation}
\label{app:small_gap}
\newcommand{\rkx}{k_x^{\operatorname{Re}}}
\newcommand{\rky}{k_y^{\operatorname{Re}}}
\newcommand{\ikx}{k_x^{\operatorname{Im}}}
\newcommand{\iky}{k_y^{\operatorname{Im}}}

In this Appendix we provide an interpretation of the exponential part of the bare propagator $e^{i\bm{k}_{j,\epsilon}(\theta_{\bm{r}})\cdot \bm{r}}$ by relating the real and imaginary parts of critical points to the normal energy dispersion. For concreteness, we consider the positive solution $\epsilon = +$ and the pocket $j=1$, and thereby we drop those labels. Generalizing to other solutions is straightforward.

We extend the domain of the normal energy dispersion to the complex plane and we separate it in its real and imaginary parts,
\begin{equation}
    \label{eq:ap-complex_ek}
    \varepsilon_{\bm{k}} = u_{\varepsilon}(\rkx, \rky, \ikx, \iky) + i u_{\varepsilon}(\rkx, \rky, \ikx, \iky),
\end{equation}
 where $\bm{k} = (k_x, k_y)$ and $k_{\nu} = k_{\nu}^{\operatorname{Re}} + i k_{\nu}^{\operatorname{Im}}$ for $\nu = x,y$. Further, we assume that $\varepsilon_{\bm{k}}$ is holomorphic and it satisfies the Cauchy-Riemann relations. Let us introduce
 \begin{equation}
     \label{eq:ap-P}
     P = \{\rkx (\theta_{\bm{r}}), \rky (\theta_{\bm{r}}), 0 ,0 \},
 \end{equation}
a point living in $\mathbb{C}^2 \cong \mathbb{R}^4$ whose projection on the real plane $ \{\rkx (\theta_{\bm{r}}), \rky (\theta_{\bm{r}})\}$ is a critical point of the normal-metal bare propagator. From the normal-metal solution we know that the critical points lie on the Fermi contours (see App.~\ref{app:nm} for details), therefore $P$ satisfies the following equations,
\begin{subequations}
    \begin{align}
        u_{\varepsilon}(P) &= 0,\\
        v_{\varepsilon}(P) &= 0,\\
        \frac{\nabla \varepsilon_{\bm{k}}}{|\nabla \varepsilon_{\bm{k}}|} \Bigr\lvert_P &= \hat{\bm{r}}.
    \end{align}
\end{subequations}
We consider the ansazt that the critical points of the superconducting propagator are a perturbation from $P$, namely
\begin{equation}
\begin{split}
    \label{eq:ap-P_sc}
    P' = \{&\rkx (\theta_{\bm{r}}) + \delta \rkx (\theta_{\bm{r}}),\\
    &\rky (\theta_{\bm{r}}) + \delta \rky (\theta_{\bm{r}}),\\
    &\delta \ikx (\theta_{\bm{r}}) ,  \delta \iky (\theta_{\bm{r}}) \}.
\end{split}
\end{equation}
This assumption will remain valid in the small-gap limit, i.e. as long as $\Delta \ll \hbar k_{\mathrm {F}} v_{\mathrm {F}}$. Naturally, $P'$ fulfills Eqs.~\eqref{eq:crit_sol_energy} and \eqref{eq:crit_sol_gradient}, therefore we have
\begin{subequations}
\label{eq:ap-eqs_p_sc}
    \begin{align}
        u_{\varepsilon}(P') &= 0,\label{eq:ap-u_p_sc}\\
        v_{\varepsilon}(P') &= \omega,\label{eq:ap-v_p_sc}\\
       g_x(P') \sin \theta_{\bm{r}} &= g_y(P') \cos \theta_{\bm{r}},\label{eq:ap-grad_p_sc}
    \end{align}
\end{subequations}
where $g_{\nu} = \partial_{k_{\nu}^{\operatorname{Re}}} u_{\varepsilon} + i \partial_{k_{\nu}^{\operatorname{Re}}} v_{\varepsilon}$ for $\nu = x, y$. Next, we develop to first order the Eqs.~\eqref{eq:ap-eqs_p_sc}. To zeroth order, the energy equations \eqref{eq:ap-u_p_sc} and \eqref{eq:ap-v_p_sc} are trivial, while the gradient equation \eqref{eq:ap-grad_p_sc} yields the following relations,
\begin{subequations}
    \begin{align}
        \partial_{\rkx} u_{\varepsilon}\Bigr\rvert_{P} \sin \theta_{\bm{r}} &= \partial_{\rky} u_{\varepsilon}\Bigr\rvert_{P} \cos \theta_{\bm{r}},\\
        \partial_{\rkx} v_{\varepsilon}\Bigr\rvert_{P} \sin \theta_{\bm{r}} &= \partial_{\rky} v_{\varepsilon}\Bigr\rvert_{P} \cos \theta_{\bm{r}}.
    \end{align}
\end{subequations}
To first order, we obtain the following condition,
\begin{equation}
    A \begin{pmatrix}
           \delta \rkx (\theta_{\bm{r}}) \\
          \delta \rky (\theta_{\bm{r}}) \\
           \delta \ikx (\theta_{\bm{r}}) \\
           \delta \iky (\theta_{\bm{r}})
         \end{pmatrix} =
         \begin{pmatrix}
           0 \\
          \omega\\
           0 \\
           0
         \end{pmatrix},
\end{equation}
where $A$ is a four by four matrix whose non-zero entries read
\begin{subequations}
    \begin{align}
        a_{11} = a_{23} &= \partial_{\rkx} u_{\varepsilon}\Bigr\rvert_{P},\\
        a_{12} = a_{24} &= \partial_{\rky} u_{\varepsilon}\Bigr\rvert_{P},\\
        a_{31} = a_{43} &= \partial^2_{\rkx \rkx} u_{\varepsilon}\Bigr\rvert_{P} \sin \theta_{\bm{r}} - \partial^2_{\rky \rkx} u_{\varepsilon}\Bigr\rvert_{P} \cos \theta_{\bm{r}},\\
        a_{32} = a_{44} &= \partial^2_{\rkx \rky} u_{\varepsilon}\Bigr\rvert_{P} \sin \theta_{\bm{r}} - \partial^2_{\rky \rky} u_{\varepsilon}\Bigr\rvert_{P} \cos \theta_{\bm{r}},
    \end{align}
\end{subequations}
where the equalities between different matrix entries follow from applying the Cauchy-Riemann relations. By expressing $\varepsilon_{\bm{k}}$ as a power series it becomes clear that the first and second order derivatives of $u_{\varepsilon}$ with respect to $\ikx$ and $\iky$ evaluated at $P$ vanish, therefore all the other matrix entries are zero.

To this order of approximation we find
\begin{equation}
    \{\delta \rkx (\theta_{\bm{r}}), \delta \rky (\theta_{\bm{r}}) \} = \{0,0\},
\end{equation}
that is, the real part of the critical points sits on the Fermi contours mimicking the normal-metal critical points. For the imaginary part which yields the exponential decay length we have
\begin{equation}
\label{eq:ap-delta_kIm}
    \delta \bm{k}^{\operatorname{Im}} \cdot \hat{\bm{r}} = \frac{\omega}{\partial_{\rkx} u_{\varepsilon}\Bigr\rvert_{P} \cos \theta_{\bm{r}} + \partial_{\rky} u_{\varepsilon}\Bigr\rvert_{P} \sin \theta_{\bm{r}}} = \frac{\omega}{|\bm{\nabla}\varepsilon_{\bm{k}}(\theta_{\bm{r}})|}.
\end{equation}
where the last equality follows from noting that the gradient of the energy dispersion is parallel to $\hat{\bm{r}}$ at the critical point.

\section{Relationship between the prefactor and the decay length of the approximate bare propagator}
\label{app:pref}

In this Appendix we provide a simple scaling argument to argue that the prefactor of the LDOS, $\Gamma (\theta_{\bm{r}})$, and the corresponding exponential decay length, $\xi (\theta_{\bm{r}})$, are always in phase. We recall that
\begin{subequations}
\begin{align}
    \xi(\theta_{\bm{r}})&\sim |\nabla \varepsilon_{\bm{k}}(\theta_{\bm{r}})|,\\
    \Gamma(\theta_{\bm{r}})&= \frac{1}{|\nabla \varepsilon_{\bm{k}}(\theta_{\bm{r}})|\sqrt{\kappa (\theta_{\bm{r}})}},
\end{align}
\end{subequations}
where it is understood that the right-hand side of the equations is evaluated at the critical point on the Fermi contours (in the small-gap approximation), and we dropped the $j,\epsilon$ labels to lighten the notation. The curvature of the energy contour is given by
\begin{equation}
\kappa = \frac{\hat{\bm{r}}_{\perp}\cdot H(\varepsilon_{\bm{k}}) \cdot \hat{\bm{r}}_{\perp}}{|\nabla \varepsilon_{\bm{k}}|},
\end{equation}
where $\hat{\bm{r}}_{\perp}$ is a unit vector perpendicular to the observation direction $\hat{\bm{r}}$ (hence perpendicular to the gradient of $\varepsilon_{\bm{k}}$ at the critical points) and $H(\varepsilon_{\bm{k}})$ is the Hessian matrix of the normal energy dispersion.

Let us consider the coordinate transformation
\begin{subequations}
\begin{align}
    x' &= \gamma x,\\
    y' &= \frac{y}{\gamma},
\end{align}
\end{subequations}
whereby we create an anisotropy of a previously isotropic dispersion, while preserving the total area of the Fermi surface. This operation decreases the curvature and increases the norm of the gradient of the energy dispersion along the $x$-direction, and vice versa along the $y$-direction.

For concreteness, let us consider a critical point sitting on the $x$ axis such that $\theta_{\bm{r}} = 0$. It follows that
\begin{subequations}
\begin{align}
    |\nabla \varepsilon_{\bm{k}}(\theta_{\bm{r}} = 0)|' &= \gamma |\nabla \varepsilon_{\bm{k}}(\theta_{\bm{r}} = 0)|,\\
    \kappa'(\theta_{\bm{r}} = 0) &= \frac{1}{\gamma^3} \kappa (\theta_{\bm{r}} = 0).
\end{align}
\end{subequations}
Putting all together we find 
\begin{subequations}
\begin{align}
    \xi'(\theta_{\bm{r}} = 0) &= \gamma \; \xi (\theta_{\bm{r}} = 0),\\ 
    \Gamma'(\theta_{\bm{r}} = 0) &= \sqrt{\gamma} \; \Gamma(\theta_{\bm{r}} = 0).
\end{align}
\end{subequations}
For the chosen coordinate transformation and observation direction,  both the prefactor and the decay length are enhanced. Obviously, in the perpendicular observation direction, one has $\gamma\rightarrow\gamma^{-1}$ so both are diminished. Therefore, as observed in Fig.~\ref{fig:single_pocket} the prefactor and the decay length are in phase as we vary the observation direction $\theta_{\bm{r}}$. The underlying reason is that even though the prefactor inversely depends on the norm of the gradient, this is compensated by an opposite behavior of the curvature.

\section{Beyond the small-gap approximation}
\label{app:beyond}
\begin{figure}
    \centering
    \includegraphics[width=0.85\columnwidth]{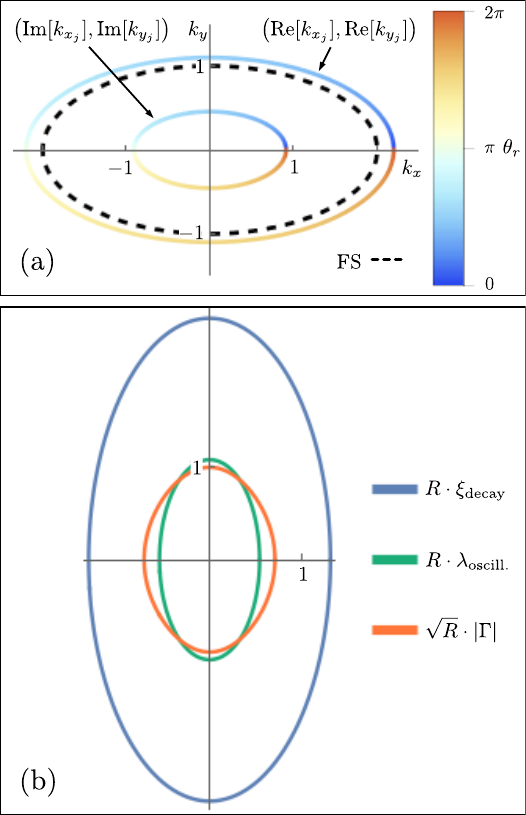}
    \caption{Saddle-point approximation beyond the small-gap limit for the ellipsoidal toy model. (a) Real and imaginary parts of the set of critical points as a function of the observation direction $\theta_{\bm{r}}$ which is color-coded. The dashed line indicates the Fermi surface. (b) Polar plot of the dimensionless quantities encoding the anisotropy of the bare propagator. The system's parameters are $\mu=\omega=\tau=1 \; \mathrm{a.u.}$ and $\alpha=0.25$}
    \label{fig:ellipse}
\end{figure}

To better understand the meaning of complex critical points it is illustrative to consider a toy model for which Eqs.~\eqref{eq:crit_sol} can be solved analytically, namely an ellipsoidal energy dispersion:
\begin{equation}
 \label{eq:ek_ellipse}
 \varepsilon_{\bm{k}} = \tau (\alpha k_x^2 + k_y^2) - \mu,
\end{equation}
where $\alpha \in (0, 1]$ controls the anisotropy (for $\alpha = 1$ the Fermi surface is a circle and in the limit $\alpha \rightarrow 0$ the Fermi surface becomes a line), $\mu$ is the chemical potential which effectively controls the ``size'' of the Fermi surface, and $\tau = \hbar k_{\mathrm{F}} v_{\mathrm{F}}$ is a trivial prefactor to adjust the units. The Fermi surface is the ellipse
\begin{equation}
 \label{eq:fs_ellipse}
 \left(\frac{x}{a}\right)^2 + \left(\frac{y}{b}\right)^2 = 1,
\end{equation}
with $a_{\mathrm{FS}} = \sqrt{\frac{\mu}{\tau \alpha}}$ and $b_{\mathrm{FS}} = \sqrt{\frac{\mu}{\tau}}$. The positive solution of Eq.\eqref{eq:crit_sol} reads
\begin{align}
\begin{split}
 \bigr(\operatorname{Re}&[k_{x_j}], \operatorname{Re}[k_{y_j}]\bigl) =\\
 &\frac{R \cos \varphi}{\sqrt{\cos^2\theta_{\mathbf{r}}+\alpha \sin^2\theta_{\mathbf{r}}}}\left(\frac{1}{\sqrt{\alpha}} \cos \theta_{\mathbf{r}}, \sqrt{\alpha} \sin \theta_{\mathbf{r}}\right),
 \end{split}
 \\[3ex]
 \begin{split}
\bigl(\operatorname{Im}&[k_{x_j}], \operatorname{Im}[k_{y_j}]\bigl) =\\
 &\frac{R \sin \varphi}{\sqrt{\cos^2\theta_{\mathbf{r}}+\alpha \sin^2\theta_{\mathbf{r}}}}\left(\frac{1}{\sqrt{\alpha}} \cos \theta_{\mathbf{r}}, \sqrt{\alpha} \sin \theta_{\mathbf{r}}\right),
\end{split}
\end{align}
where $R = \left(\frac{\mu^2+\omega^2}{\tau^2}\right)^{1/4}$, and $\varphi = \frac{1}{2}\arctan{\frac{\omega}{\mu}}$. The negative solution of Eq.~\eqref{eq:crit_sol} is analogous up to an overall minus sign for the real part, in agreement with the relationship discussed at the end of Section \ref{sec:results} for even energy dispersions. The real and imaginary parts of the critical points lie on ellipses [Fig.~\ref{fig:ellipse} (a)] such that
\begin{align}
 a_{\mathrm{Re}} = \frac{R\cos \varphi}{\sqrt{\alpha}}, \qquad b_{\mathrm{Re}} = R \cos \varphi,\\
 a_{\mathrm{Im}} = \frac{R\sin \varphi}{\sqrt{\alpha}}, \qquad b_{\mathrm{Im}} = R \sin \varphi.
 \end{align}
To first order in $\omega$ we have $(a_{\mathrm{Re}}, b_{\mathrm{Re}}) \approx (a_{\mathrm{FS}}, b_{\mathrm{FS}})$ and $(a_{\mathrm{Im}}, b_{\mathrm{Im}}) \approx \frac{\omega}{2\mu}(a_{\mathrm{FS}}, b_{\mathrm{FS}})$. As we discussed in Section \ref{subsec:small_gap}, in the small-gap limit the real part of the critical points collapses to the Fermi surface, while the imaginary counterpart is linear in the superconducting parameter. The relevant quantities characterizing the anisotropy of the bare propagator follow:
\begin{align}
\begin{split}
    \xi_{\mathrm{decay}}(\theta_{\bm{r}}) &\equiv \frac{1}{\operatorname{Im}[\bm{k}(\theta_{\bm{r}})]\cdot \hat{\bm{r}}}\\
    &= \frac{1}{R\sin \varphi}\sqrt{\frac{\alpha}{\cos^2\theta_{\bm{r}} + \alpha \sin^2\theta_{\bm{r}}}},
\end{split}
\\[3ex]
\begin{split}
    \lambda_{\mathrm{oscill.}}(\theta_{\bm{r}}) &\equiv \frac{1}{\operatorname{Re}[\bm{k}(\theta_{\bm{r}})]\cdot \hat{\bm{r}}}\\
    &= \frac{1}{R\cos \varphi}\sqrt{\frac{\alpha}{\cos^2\theta_{\bm{r}} + \alpha \sin^2\theta_{\bm{r}}}},
\end{split}
\\[3ex]
\begin{split}
    \Gamma(\theta_{\bm{r}}) &\equiv \frac{1}{|\bm{\nabla}\varepsilon_{\bm{k}(\theta_{\bm{r}})}|\sqrt{\kappa_{\bm{k}(\theta_{\bm{r}})}}} = \\
    &\frac{1}{2 \sqrt{R} (\cos^2\theta_{\bm{r}} + \alpha \sin^2\theta_{\bm{r}})^{1/4}}e^{-i\frac{\varphi}{2}}.
\end{split}
\end{align}

As shown in Fig.~\ref{fig:ellipse} (b) the propagator is enhanced along directions perpendicular to flatter sections of the Fermi surface and the decay length $\xi_{\mathrm{decay}}$ is in phase.

\begin{acknowledgments}
We thank T. Cren, F. Massee, A. Palacio-Morales, J. C. S. Davis and M. Aprili for fruitful discussions.
\end{acknowledgments}

\bibliography{bibliography.bib}

\end{document}